\documentstyle[preprint,aps]{revtex}
\tightenlines
\begin{document}
\draft
\preprint{\vbox{
\hbox{IFT-P.24/98}
\hbox{hep-ph/9804422}
\hbox{April 1998}
}}
\title{ Concerning CP violation in 331 models~\footnote{Talk given by O. Ravinez
at the 7th Workshop on Particles and Fields, Morelia, Mexico, 21-27 Nov 1997.}} 
\author{ J. C. Montero, V. Pleitez and O. Ravinez }  
\address{
Instituto de F\'\i sica Te\'orica\\
Universidade Estadual Paulista\\
Rua Pamplona 145\\ 
01405-900-- S\~ao Paulo, SP\\Brazil} 
\maketitle
\begin{abstract}
We consider the implementation of CP violation in the context of 331 models.
In particular we treat a model where only three scalar triplets are needed in
order to give all fermions a mass while keeping neutrino massless. In this 
case all $CP$ violation is provided by the scalar sector.  
\end{abstract}

In spite of the great efforts of theoreticians and experimentalists, 
the origin and the smallness of $CP$ violation remains an open question. 
In the context of the
standard electroweak model~\cite{sm} the $CP$ symmetry is violated in the 
complex Yukawa couplings~\cite{km}. Although this is an interesting feature of 
the model it leaves open the question of why $CP$ is so feebly violated. 
Since the works of Lee and Weinberg it has been known that in 
renormalizable gauge theories the violation of $CP$ has the right strength
if it occurs through the exchange of a Higgs boson of mass $M_H$~\cite{tdlsw}, 
{\it i. e.,} it is proportional to $G_Fm^2_f/M^2_H$, where $m_f$ is the fermion 
mass. Since then, there have been
many realizations of that mechanism in extensions of the electroweak standard 
model~\cite{cp}.
Recently, it has been proposed models with the electroweak gauge group
being $SU(3)_L\otimes U(1)_N$ instead of the usual $SU(2)_L\otimes U(1)_Y$
~\cite{ppf,pt}. 
One interesting feature of these sort of models is that the anomalies cancel 
only when all three families are taken together. Although neutrinos remain
massless there is lepton-flavor violation in the interactions with doubly
charged scalar and vector bosons which are present in the model.
Hence, it is possible to have $CP$ violation in that mixing 
matrix~\cite{liung}. Another possibility is purely spontaneous $CP$ violation
through complex value for the vacuum expectation values (VEVs) for the neutral
scalars. This however, only happens in the model with three triplets and one 
sextet~\cite{laplata1}.

Let us consider a model with 331 symmetry with exotic heavy leptons. The
three leptons generations belong to $({\bf1},{\bf3},0)$ representation.
It means $(\nu_l,l^-,E_l^+)^T$, for $l=e,\mu,\tau$. The scalar content of the
model necessary to give masses to all fermions is
\begin{equation}
{\bf \chi }=\left( 
\begin{array}{c}
\chi ^{-} \\ 
\chi ^{--} \\ 
\chi ^0
\end{array}
\right) \sim \left( {\bf 3},{\bf -1}\right), \quad {\bf \rho }=\left( 
\begin{array}{c}
\rho ^{+} \\ 
\rho ^0 \\ 
\rho ^{++}
\end{array}
\right) \sim \left( {\bf 3},{\bf 1}\right), \quad 
{\bf \eta }=\left( 
\begin{array}{c}
\eta ^0 \\ 
\eta _1^{-} \\ 
\eta _2^{+}
\end{array}
\right) \sim \left( {\bf 3},{\bf 0}\right).
\label{e1}
\end{equation}
As we said before, the neutrinos are optionally massless if we do not 
introduce the right-handed components.

We allow VEVs being complex numbers {\it i.e.,} $v_a=
\vert v_a\vert\exp(i\theta_a)$, where $a=\eta,\rho$ and $\chi$.
However, it is not enough to implement $CP$ violation.
The minimization of the potential implies the conditions 
${\rm Im}(v_\eta v_\rho v_\chi)=0$ and, since we can choose always two 
VEVs being real because of the $SU(3)$ symmetry, it means that no
phase at all survive in the potential minimum~\cite{laplata1}.
However, if we allow beside the complex VEVs, the trilinear term in the 
potential $\alpha \epsilon_{ijk}\eta_i\rho_k\chi_k+H.c.$ (where $i,j,k$ are 
$SU(3)$ indices) with the complex constant 
$\alpha=\vert \alpha\vert e^{i\theta_\alpha}$, the minimization 
of the potential in this case implies ${\rm Im}(\alpha v_\eta v_\rho v_\chi)=0$
and the relative phase, say, among $\alpha$ and $v_\chi$ will survive in
the Lagrangian density. Hence, there are explicit $CP$ violation in the 
Lagrangian. (Notice however, that $CP$ violation also requires complex VEV's.
We will assume real Yukawa couplings too.)

The lepton are assigned to the following representations: 
\begin{equation}
\Psi_{aL}=\left(\begin{array}{c}
\nu_{l_a} \\  l'^-_a\\ E'^+_a
\end{array}\right)_L \sim ({\bf3},0);\quad l'^-_{aR}\sim({\bf1},-1),\;
E'^-_{aR}\sim({\bf1},+1),\;\;a = e, \, \mu , \, \tau.
\label{lep1}
\end{equation}
It is possible to absorb all phases in the leptonic mass matrix so that
the symmetry eigenstates (primed fields) are related to the mass eigenstates
(unprimed fields) thorough orthogonal matrices~\cite{cb}
\begin{equation}
l'_{aL}={\cal O}^e_{Lai}l_{iL},\quad
l'_{aR}={\cal O}^e_{Rai}l_{iR},\quad
E'_{aL}={\cal O}^E_{Lai}l_{iL},\quad
l'_{aR}={\cal O}^E_{Rai}l_{iR},\quad i=1,2,3.
\label{redef}
\end{equation}

It is always possible to choose the $CP$ violation to
occur only through the exchange of singly and doubly charged scalar fields. 
(The $CP$ violation in the neutral Higgs sector is transformed away by a 
redefinition of the right-handed components of the lepton fields, say 
$e_{iR}\to \exp(i\theta_\rho)e_{iR}$ and $E_{iR}\to \exp(i\theta_\chi)E_{iR}$.)

The Yukawa couplings of leptons with the doubly charged scalars are
\begin{equation}
{\cal L}_Y=\frac{\sqrt{2}}{\vert v_\rho\vert}\,
\bar{E}_{L}({\cal O}^E_L)^T{\cal O}^e_LM^e\,e_L
\rho^{++} +
\frac{\sqrt{2}}{\vert v_\chi\vert}e^{-i\theta_\chi}
\bar{e}_L({\cal O}^e_L)^T{\cal O}^E_LM^E \,E_LL\chi^{++}+H.c.,
\label{yuka}
\end{equation}
where we have used $\Gamma^e={\cal O}^e_LM^e{\cal O}^{eT}_R$,
$\Gamma^E={\cal O}^E_LM^E{\cal O}^{ET}_R$ with $M^e=
{\rm diag}(m_e,m_\mu,m_\tau)$; $M^E=
{\rm diag}(m_{E_1},m_{E_2},m_{E_3})$ (where $\sqrt2\Gamma^e/v_\rho$ and
$\sqrt2 \Gamma^e/v_\chi$ are the arbitrary Yukawa dimensionless couplings). 
In Eq.~(\ref{yuka}) the scalar fields are still symmetry eigenstates.
In fact, there are one Goldstone boson, $G^{++}$ and a physical one, $X^{++}$,
we denote its mass by $m_X$. (In this model there is not lepton-number  
violation in the interactions with the neutral scalars.)
We have verified that it is not possible to absorb all phases in the complete 
Lagrangian density. We have then $CP$ violation through the exchange of
physical scalars. 

The standard model prediction for the EDM of the electron is rather small, of 
the order of magnitude of $2\times 10^{-38}$ e~cm~\cite{hoo}. On the other 
hand, the experimental upper limit is  $\leq 4\times 10^{-27}$ e~cm~\cite{ele}. 
Hence, it is interesting that if a large value for the electron EDM (and 
other elementary particles) is found, it would indicate new 
physics beyond the standard model. If neutrino remains 
massless in the standard model the contribution to electric dipole moment will 
arise at the three loop level~\cite{hoo} (or, at the two level in other
models with massless neutrinos~\cite{chao}). In our case we have

\begin{equation}
d_e=-\frac{em_e}{64\pi^2m^2_X}
\,\sqrt{2}M^2_WG_F\;{\bf O}_{ee}\;
\sin(2\theta_\alpha),
\label{emde}
\end{equation}
where we have defined
\begin{equation}
{\bf O}_{ee}=\sum_j
\left[\left({\cal O}^e_L\right)^T {\cal O}^E_L\right]^2_{ej}\,
\;\frac{4m^2_{E_j}}{M^2_U}\;[F_+(m_{E_j})+F_-(m_{E_j})],
\label{oo}
\end{equation}
with
\begin{equation}
F_\pm(m_{E_j})\!\!=\!\!-\frac{m^2_X}{2m_e^2}\ln\frac{m^2_X}{m^2_{E_j}}
+\frac{m^2_X}{2m_e^2}(m^2_X\pm m^2_e-m^2_{E_j})\Delta^{-1}\,
\ln\left[ \frac{m^2_{E_j}+m^2_X-m^2_e+\Delta}
{m^2_{E_j}+m^2_X-m^2_e-\Delta}\right],
\label{fs}
\end{equation}
and
\begin{equation}
\Delta^2=(m^2_X+m^2_{E_i}-m^2_e)(m^2_X-m^2_{E_i}-m^2_e).
\label{delta}
\end{equation}
In writing Eq.~(\ref{emde}) we have used 
$M^2_U=(g^2/4)(\vert v_\chi\vert^2+\vert v_\rho\vert^2)$ and 
$G_F/\sqrt2=g^2/8M^2_W$ where $M_U$ is the mass of the doubly charged vector 
boson that is present in the model. We have chosen $\theta_\eta=
\theta_\rho=0$, and $\theta_\chi=-\theta_\alpha$. 

For nondegenerate heavy leptons the mixing angles remain in Eq.~(\ref{oo}).
For instance, the contribution of $E_1$, using $m_{E_1}=50$ GeV, $m_\chi=100$ 
GeV~\cite{pdg}, we obtain 
\begin{equation}
d_e\approx -\left(\frac{M^2_W}{M^2_U}\right)\,
\left[\left({\cal O}^e_L\right)^T {\cal O}^E_L\right]^2_{e1}\,
 \sin(2\theta_\alpha)\,\times 10^{-17}\;{\rm e \, cm}.
\label{edme}
\end{equation}
Assuming $M_U=300$ GeV and that the factor with the mixing angles (including 
$\sin(2\theta_\alpha)$ is $\approx10^{-8}$
we obtain $d_e\approx 10^{-27}$ e~cm, which
is compatible with the experimental upper limit of $10^{-27}$ e~cm~\cite{ele}.

For the muon the experimental EDM's upper limit
are of the order of $<10^{-19}$ e~cm~\cite{muon}. It means a constraint in 
$\left[\left({\cal O}^e_L\right)^T {\cal O}^E_L\right]^2_{\mu2}\leq 1$.
For the tau lepton a limit of $10^{-17}$ e~cm is derived from 
$\Gamma(Z\to \tau^+\tau^-)$~\cite{escribano}. In the present model 
the EDM of the tau lepton is at least of the order of $10^{-19}$ e~cm.

In this model there is not rare decays such as $\mu\to 3e$ at tree level. 
However, the same loop diagrams that contribute for the EDM of the leptons
imply also magnetic and electric-moment transitions, like $\mu\to e\gamma$.
This transitions will constrain the matrix elements 
$\left[\left({\cal O}^e_L\right)^T {\cal O}^E_L\right]^2_{\mu1} $ only.

In the quark sector we have contributions involving the exchange of 
one simple charged and one doubly charged scalars in the box diagrams
that contribute to the $\varepsilon$ and $\varepsilon'$ parameters of the
neutral Kaon system. There are also contributions to the electric dipole moment
of the neutron. These issues will be published elsewhere~\cite{cp3}.

With three triplets and one sextet which are needed in the model of 
Ref.~\cite{ppf} it is possible to have truly spontaneous violation of the $CP$ 
symmetry. In this case, the minimization condition of the scalar potential 
implies ${\rm Im}(v_Sv_\eta v_\rho v_\chi)=0$~\cite{laplata1},
with $v_S$ the VEV of one of the neutral component of the sextet which gives 
mass to the charged leptons. The phenomenology of this model has been studied 
in Ref.~\cite{laplata2}.

\end{document}